# Comment on "Performance of a spin-based insulated gate field effect transistor" [cond-mat/0603260]

S. Bandyopadhyay

*Department of Electrical and Computer Engineering, Virginia Commonwealth University, Richmond, VA 23284, USA*

M. Cahay

*ECECS Department, University of Cincinnati, Cincinnati, OH 45221, USA*

In a recent e-print [1], Hall and Flatte (HF) claim that a particular spin based field effect transistor (SPINFET), which they have analyzed, will have a lower threshold voltage, lower switching energy and lower leakage current than a comparable metal oxide semiconductor field effect transistor (MOSFET). This contradicts ref. [2] which found that the opposite is generally true. Here, we show that all three claims of HF are invalid and the conclusion of ref. [2] still stands.

The first claim of HF is that their SPINFET has a lower threshold (or switching) voltage than a comparable MOSFET. For a meaningful comparison, one needs to compare like entities. Since the HF structure can be used either as a SPINFET or a MOSFET, what will be meaningful is to find out *in which mode* the switching voltage will be smaller. HF calculate that when used in the SPINFET mode, the switching voltage for a 20 nm thick InAs channel is 100 mV for a carrier concentration of 2 x $10^{12}$ cm$^{-2}$ [1] and 140 mV for a carrier concentration of 2 x $10^{11}$ cm$^{-2}$ [3]. When used in the MOSFET mode, the switching voltage will be simply the voltage required to accumulate or deplete carriers in the channel, depending on whether it is an enhancement or depletion type device. If a single subband is occupied, then this voltage is $E_F/e = \pi\hbar^2 n_s/(m^* e)$, where $E_F$ is the Fermi energy, $e$ is the electron charge, $n_s$ is the surface carrier concentration and $m^*$ is the carrier effective mass. Thus, for an InAs well, the switching voltages in the MOSFET mode will be 157 mV and 15.7 mV, respectively, for the two carrier concentrations that HF considered. Therefore, a higher carrier concentration favors the HF device and a lower one favors the MOSFET. The MOSFET *could always win by lowering the carrier concentration in the channel*. Since the switching energy is proportional to the square of the switching voltage, the same is true of the switching energy. Consequently, the HF claim that their device has a lower threshold and a lower switching energy is not generally admissible.

The comparison is even worse when it comes to leakage current. Here the HF device has a serious problem which may make it entirely useless as a transistor. This device is fashioned after the device of ref. [4]. The basic idea there is to inject 100% spin polarized electrons into the channel of a transistor from an ideal magnetized ferromagnetic source. If the spin relaxation time in the channel ($T_1$) is much longer than the transit time $\tau_{transit}$, then the injected electrons will reach the drain with their spin polarizations mostly intact. The ferromagnetic drain is magnetized anti-parallel to the source so that it blocks these carriers from transmitting and the source-to-drain current is nearly zero. This is the OFF state of the device. Now if the gate voltage alters the spin orbit interaction in the channel and makes $T_1 << \tau_{transit}$, then spins will flip in the channel and the flipped ones can transmit through the drain, resulting in larger current. This is the ON state. It is obvious that carriers reaching the drain in the ON state will have *at best* equal spin-up and spin-down populations so that a maximum of 50% will transmit, while ~ 0% transmits in the OFF state. Thus, the on-off ratio of the current can ideally reach $0.5/0.0 = \infty$, *but only if the source and drain act as ideal spin injectors and detectors that inject and transmit only majority spins with 100% efficiency*.

Ref. [1] has implicitly assumed 100% spin injection and detection efficiency in all its calculations, which is unrealistic. There are no known 100% electrical spin injectors and detectors. Even after more than a decade of research, the largest demonstrated electrical spin injection efficiency is 70% using permanent ferromagnets [5] and 90% using non-permanent ones at a temperature of 4.2 K [6]. If we assume the former value, then 15% of the carriers injected by the source will always have minority spins. If we further assume (generously) that the drain is ideal and transmits all of them while blocking all the opposite spins, then the leakage current in the OFF state is carried by at least 15% of the carriers, while the on-current is carried by at most 50% of the carriers. Therefore, the real on-off ratio of the current is no more than 0.5/0.15 = 3.3. That immediately disqualifies such a device from being a serious contender in any mainstream application. If we can increase the injection efficiency to 90%, the on-off ratio increases to a maximum of 10, which is still inadequate. To reach an on-off ratio of 1000 (required for fault-tolerant digital circuits), the spin injection (and detection) efficiency has to reach 99.9%! Therefore, realistically, the on-off ratio will be low, probably < 10 at room temperature, which precludes all mainstream applications. Increasing the carrier concentration to reduce the threshold voltage will further reduce the spin injection efficiency and exacerbate the problem.

The low on-off ratio causes high error rates in digital applications, and also high leakage current in the OFF-state. Assuming 70% spin-injection efficiency, the leakage current in the OFF-state is at least one-third of the on-current. Based on Fig. 3(a) of ref. [1], the leakage current will then be larger than 3 µA/µm. This value is $10^5$ times larger than that calculated by HF (see Table I of ref. [1]) who assumed a 100% injection/detection efficiency. With a leakage current this high, the resulting standby power dissipation will be unacceptable. Thus, instead of reducing overall power dissipation, the HF device will increase power dissipation tremendously.

We therefore conclude that the HF device has no advantage over a MOSFET. Rather, it has severe shortcomings.